\documentclass[11pt,preprint]{aastex}

\begin{document}

\pagestyle{myheadings}
\markright{{\it Conference Highlights}}

\begin{titlepage}

\title{\Large{\bf Radio Astronomy in LSST Era}\footnote{Workshop was
held in Charlottesville, \hbox{VA}, on~2013 May~6--8}}
\noindent
{\it \phantom{0}}
\\%
{\it T.~Joseph~W.~Lazio} 
\noindent
({\it Jet Propulsion Laboratory, California Institute of Technology})
\\%
\noindent
{\it A.~Kimball} 
\noindent
({\it National Radio Astronomy Observatory and CSIRO Astronomy \& Space
	Science})
\\%
\noindent
{\it A.~J.~Barger}
\noindent
({\it University of Wisconsin-Madison})
\\%
\noindent
{\it W.~N.~Brandt}
\noindent
({\it The Pennsylvania State University})
\\%
\noindent
{\it S.~Chatterjee}
\noindent
({\it Cornell University})
\\%
\noindent
{\it T.~E.~Clarke}
\noindent
({\it Naval Research Laboratory})
\\%
\noindent
{\it J.~J.~Condon}
\noindent
({\it National Radio Astronomy Observatory})
\\%
\noindent
{\it Robert L.~Dickman}
\noindent
({\it National Radio Astronomy Observatory})
\\%
\noindent
{\it M.~T.~Hunyh}
\noindent
({\it International Centre for Radio Astronomy Research})
\\%
\noindent
{\it Matt J.~Jarvis}
\noindent
({\it University of Oxford and University of the Western Cape})
\\%
\noindent
{\it Mario Juri{\'c}}
\noindent
({\it LSST Corporation})
\\%
\noindent
{\it N.~E.~Kassim}
\noindent
({\it Naval Research Laboratory})
\\%
\noindent
{\it S.~T.~Myers}
\noindent
({\it National Radio Astronomy Observatory})
\\%
\noindent
{\it Samaya Nissanke}
\noindent
({\it California Institute of Technology})
\\%
\noindent
{\it Rachel Osten}
\noindent
({\it Space Telescope Science Institute})
\\%
\noindent
{\it B.~A. Zauderer}
\noindent
({\it NSF Astronomy \& Astrophysics Fellow, Harvard University})

\begin{figure*}[bh]
\centering
\includegraphics[width=0.67\textwidth]{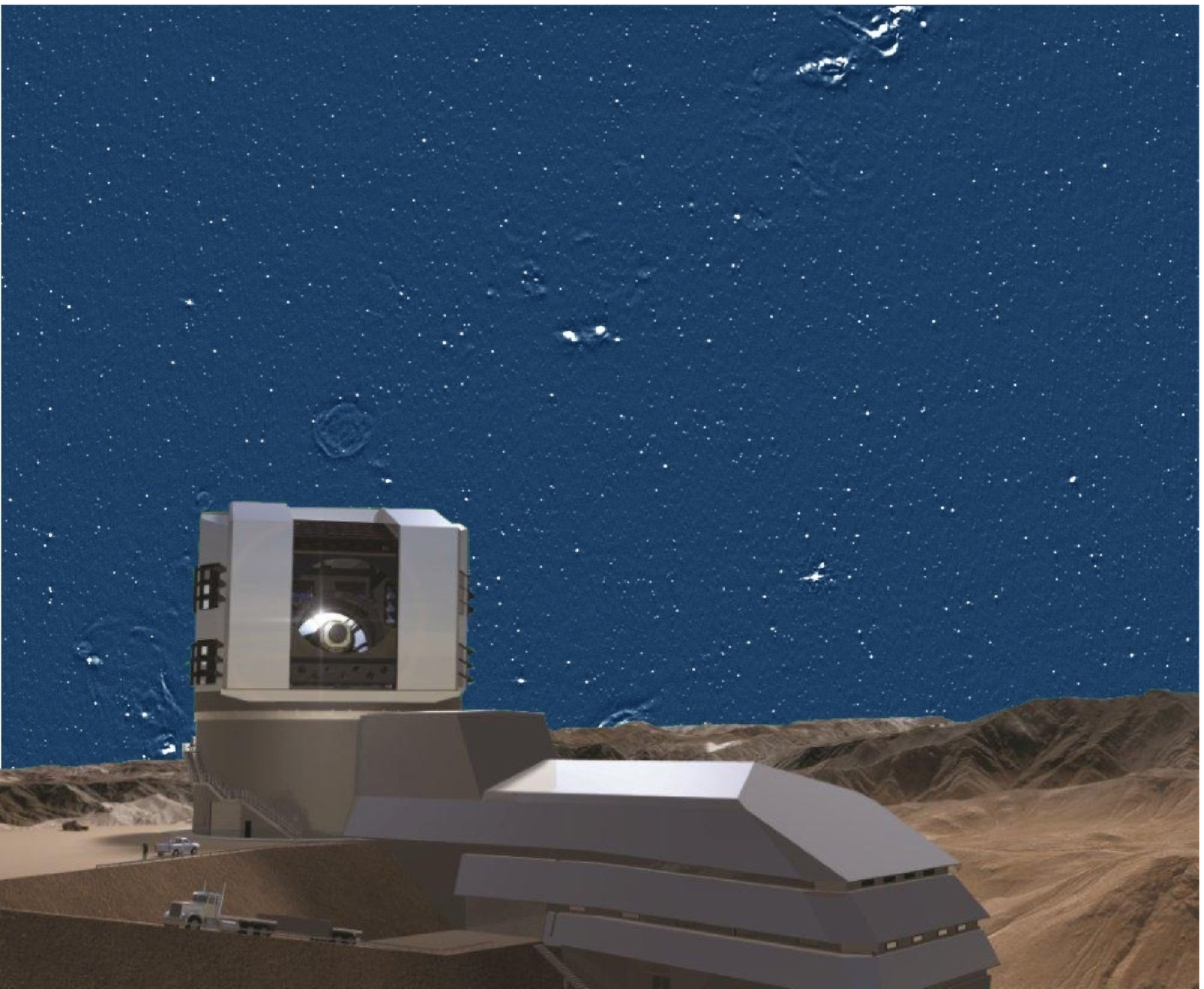}\\
\end{figure*}
\end{titlepage}

\begin{abstract}
A community meeting on the topic of ``Radio Astronomy in the LSST
Era'' was hosted by the National Radio Astronomy Observatory in
Charlottesville, \hbox{VA} (2013 May~6--8).  The focus of the workshop
was on time domain radio astronomy and sky surveys.  For the time
domain, the extent to which radio and visible wavelength observations
are required to understand several classes of transients was stressed,
but there are also classes of radio transients for which no visible
wavelength counterpart is yet known, providing an opportunity for
discovery.  From the LSST perspective, the LSST is expected to
generate as many as 1~million alerts nightly, which will require even
more selective specification and identification of the classes and
characteristics of transients that can warrant follow up, at radio or
any wavelength.  The LSST will also conduct a deep survey of the sky,
producing a catalog expected to contain over 38~billion objects in it.
Deep radio wavelength sky surveys will also be conducted on a
comparable time scale, and radio and visible wavelength observations
are part of the multi-wavelength approach needed to classify and
understand these objects.  Radio wavelengths are valuable because they
are unaffected by dust obscuration and, for galaxies, contain
contributions both from star formation and from active galactic
nuclei.  The workshop touched on several other topics, on which there
was consensus including the placement of other LSST ``Deep Drilling
Fields,'' inter-operability of software tools, and the challenge of
filtering and exploiting the LSST data stream.  There were also topics
for which there was insufficient time for full discussion or for which
no consensus was reached, which included the procedures for following
up on LSST observations and the nature for future support of
researchers desiring to use LSST data products.
\end{abstract}


\clearpage

\section{Introduction}\label{sec:intro}

By the middle of next decade, the Large Synoptic Survey Telescope
(LSST) will be in the midst of a decade-long sky survey, en route to
producing a deep, multi-color view of the sky and generating
potentially a million alerts nightly.
Radio wavelength observations will provide independent, complementary
views of the sky, and they will be crucial for full exploitation of
the LSST data products.  A community workshop, hosted by the National
Radio Astronomy Observatory in Charlottesville, \hbox{VA} (2013
May~6--8), was held to explore the science themes of time domain radio
astronomy and the radio components of multi-wavelength sky surveys,
with the aim of identifying emerging scientific and technical
capabilities needed for conducting observations in these areas.  The
focus of this workshop was primarily on centimeter- and
meter-wavelength observations ($\sim 30$~MHz--50~GHz).
This
document is intended to capture the key conclusions and
recommendations of that community workshop.\footnote{
The full science program and presentations are available at 
\texttt{https://science.nrao.edu/science/event/RALSST2013}.
}
Section~\ref{sec:landscape} (and Appendix~\ref{app:landscape})
summarizes the likely radio astronomy landscape by the middle of next
decade, Section~\ref{sec:time} discusses the complementary nature of
radio astronomical observations and LSST observations,
Section~\ref{sec:survey} describes radio wavelength surveys and their
complementary nature to the LSST survey, and Section~\ref{sec:future}
summarizes other topics of discussion at the workshop.
Appendix~\ref{app:lsst} summarizes LSST data products,
Appendix~\ref{app:surveys} describes potential future radio surveys,
and Appendix~\ref{app:who} lists the workshop participants.

\section{Radio Astronomy Landscape}\label{sec:landscape}

Observations at centimeter- to meter wavelengths have been responsible
for the discovery of many of the objects and much of the phenomena
studied in modern astronomy and have resulted in the awarding of three
Nobel Prizes in Physics.  Discoveries include the cosmic microwave background
(CMB), quasars, pulsars, indirect evidence for gravitational waves,
astrophysical masers, cosmic magnetic fields, non-thermal emission
mechanisms (specifically synchrotron radiation), and the jets from
black holes and other objects.
Moreover, the use of aperture synthesis techniques---also recognized by the Nobel
committee---has enabled observations at radio wavelengths to reach
unprecedented levels of imaging resolution and astrometric precision
not attained in any other wavelength band, providing the fuel for further discovery.

Technological developments over the latter half of the
$20^{\mathrm{th}}$ Century, often stimulated by commercial
considerations, offer a path to substantial improvements in future
radio astronomical instrumentation.  Among the range of improvements are
mass production of centimeter-wavelength antennas, fiber optics for
the transmission of large volumes of data, high-speed digital signal
processing hardware for the analysis of the signals, and computational
improvements leading to massive processing and storage.  Applying
these new technologies to radio astronomy can open up an enormous
expanded volume of discovery space.  A vigorous international program
of upgrading existing radio telescopes and constructing new ones has
been stimulated by the potential of applying these technologies to
radio astronomy, a process that is likely to continue at least through
the current decade.

While it is difficult to capture the full range of international
activity, it spans much of the available parameter space for radio
wavelength observations.  Various radio telescopes now allow access to
frequencies near the ionospheric cutoff ($\sim 20$~MHz) to as high as the significant tropospheric oxygen absorption lines ($\sim 50~$~GHz).
Filled aperture radio telescopes are in operation providing significant
surface brightness sensitivity, while interferometers, using the
technique of very long baseline interferometry (VLBI), routinely obtain
sub-milliarcsecond resolutions.  With modern digital signal
processing, sub-microsecond time resolution has been obtained, and, in
conjunction with data archives, decade-long light curves of some
objects are being produced.  Full polarization measurements of the
electric field can also be obtained routinely.  Even more significant
expansions in one or more directions in parameter space will be
enabled with future instruments, including the Hydrogen Epoch of
Reionization Array \citep[\hbox{HERA}, ][]{hera}; the Square Kilometre
Array \citep[\hbox{SKA}, ][]{ska100,dtmmlc13}; and a high frequency
long-baseline standalone U.{}S.\ array, possibly a component of the
\hbox{SKA}, grown from the Karl G.~Jansky Very Large Array (JVLA) and
Very Long Baseline Array (VLBA).\footnote{
A concept for such an array, called the North America Array, was
presented in a white paper for the 2010 Astronomy \& Astrophysics
Decacadal Survey, \texttt{http://www.nrao.edu/A2010/rfi/PPP-NAA-edited.pdf}.
}

Appendix~\ref{app:landscape} provides more quantitative information
about radio telescopes likely to become operational \emph{before} or
during the LSST era.  In the spirit of this workshop, these summaries
of telescope capabilities are intended to provoke thinking about the
kinds of radio wavelength observations that could be conducted both
\emph{now} and during the LSST era.

\section{Time Domain Astronomy}\label{sec:time}


Time domain observations at radio wavelengths have a long history,
with the initial theoretical prediction for the existence of neutron
stars \citep{bz34} being verified by their discovery at a radio
telescope that had a fortuitous combination of a wide field of view
and high time resolution \citep{hbpsc68}.  Since then, it has been
recognized that variable and transient emissions---bursts, flares,
pulses---mark compact sources or the locations of explosive or dynamic
events.  As such, radio transient sources offer insight into a variety
of fundamental physical and astrophysical questions such as the
cosmological star formation history (through observations of
supernovae and gamma-ray bursts), the nature of strong gravity
(through observations of radio pulsars), and
mechanisms for efficient particle acceleration.  Considerable recent
excitement has been generated by the report of millisecond radio
bursts having properties consistent with a cosmological population of
sources \citep{tsb+13}, suggesting that the radio sky may harbor
other, yet-to-be-discovered populations of transient sources as well.
The wide field of view and cadence of the LSST suggest that it may
have a similar discovery potential as some of the early radio
telescopes.

Table~\ref{tab:transients} summarizes various classes of known radio
transients, many of which are likely to have LSST counterparts.  
Many of these classes have visible wavelength counterparts, and there
are many classes of transients that are discovered first at shorter
wavelengths before being observed at radio wavelengths.  Consequently,
fully exploiting the LSST data stream, particularly to include
understanding new classes of sources that LSST might detect, is likely
to require some amount of follow-up radio observations.  Moreover,
is possible that a new class of radio transients might be discovered
for which LSST observations would be useful in characterizing the
class or determining what the visible wavelength counterparts (if any) are.

\begin{deluxetable}{lp{0.15\textwidth}p{0.15\textwidth}p{0.15\textwidth}c}
\tablecaption{Classes of Radio Transients\tablenotemark{a}\label{tab:transients}}
\tablewidth{0pc}
\tabletypesize{\scriptsize}
\tablehead{%
 \colhead{Class} & \colhead{Object} & \colhead{Timescale} &
 \colhead{$\Delta t_{\mathrm{opt}}$\tablenotemark{b}} & \colhead{Frequency Range}
}
\startdata

extragalactic incoherent &
SNe, GRBs, TDEs &
 	tens of minutes--years &
	lags by minutes--months, cascading in frequency &
	$\sim 0.1$--50~GHz
	\\
                         &
AGN   & 
	tens of minutes--years &
	lags &
	$\sim 0.5$--50~GHz
	\\

                         &
gravitational wave event &
 	tens of minutes?--years? &
	lags(?) by weeks--years, cascading in frequency &
	$\sim 0.1$--50~GHz
	\\

\noalign{\hrule} \\

extragalactic coherent  &
fast radio burst? & 
	sub-second &
	unknown &
	1.4~GHz\tablenotemark{c}
	\\
			&
gravitational wave event? &
	sub-second? &
	unknown &
	$\lesssim 1\,\mathrm{GHz}$? 
	\\

\noalign{\hrule} \\

Galactic coherent & 
circumstellar, interstellar masers &
	?? &
	(not applicable) &
	$\sim 1.6$--22~GHz
	 \\

				& 
neutron stars	& 
	sub-second & 
	simultaneous, if present &
	$\sim 0.1$--40~GHz
	\\

				&
sub-stellar objects & 
	sub-second--hours & 
	unknown &
	0.01--10~GHz
	\\

\noalign{\hrule} \\
Galactic incoherent & 
synchrotron flares, late-type stars, novae, colliding stellar winds &
	minutes--hours & 
	lags by minutes & 
	$\sim 1$--200~GHz
	\\

\noalign{\hrule} \\

unknown &
``Hyman bursters''  &
	minutes &
	unknown & 
	$\lesssim 1$~GHz
	\\

\noalign{\hrule} \\

propagation effects & 
	affects pulsars, compact extragalactic sources &
	minutes--days (pulsars), hours--years (AGN) & 
	(not applicable) &
	$\lesssim 5$~GHz
	\\
\enddata
\tablenotetext{a}{See text for references.}
\tablenotetext{b}{The amount by which the radio emission leads or lags
	the visible wavelength emission.}
\tablenotetext{c}{To date, the only published observations of fast
	radio bursts are at~1.4~GHz.}
\end{deluxetable}

The intrinsic transient radio emission or variability from
astronomical objects can be broadly grouped into four classes,
depending upon their location and emission mechanism.  
Radio transients can be located in the Galaxy or can be extragalactic, and
their emission mechanisms can be either \emph{incoherent} (e.g.,
synchrotron) or \emph{coherent} (e.g., masers).  For incoherent
processes, particles radiate independently (luminosity $L \propto N$,
for~$N$ radiating particles), which tends to favor centimeter (and
shorter) wavelengths by virtue of the Rayleigh-Jeans approximation.
For coherent processes, the emission occurs in phase, which can result
from stimulated emission, such as occurs in masers, or from collective
processes in which the radiating particles are ``bunched'' within a
wavelength (with $L \propto N^2$).  
The wavelength of maser emission depends upon the transitions (e.g.,
\hbox{OH}, H${}_2$O), but collective processes tend to favor meter
wavelengths because the volume within which particles can radiate in phase
scales approximately as~$\lambda^3$.  These classes are not
necessarily exclusive, as some kinds of objects 
could potentially display
incoherent and coherent emissions at different times.

Radio signals are also affected by their propagation through a plasma,
e.g., interplanetary medium, interstellar medium, or intergalactic
medium \citep{r90}.  These propagation effects tend to be strongly
wavelength dependent, and are most significant for compact sources,
but can produce considerable apparent time variability.  In extreme
cases, the propagation-induced variability
can exceed 100\%, producing brightness changes that far exceed any
intrinsic variability.

\subsection{Transients at Radio Wavelengths}\label{sec:radiotransients}

Many examples of coherent and incoherent Galactic radio transients are
well known, as well as incoherent extragalactic sources, as shown in
Table~\ref{tab:transients}. However, only recently have examples
of coherent radio transients that are potentially located at extragalactic
distances been discovered.\footnote{
While the pulsars known in the Magellanic Clouds are examples of
coherent extragalactic sources, we
consider them as extensions of a known Galactic population, rather
than examples of a coherently emitting radio source detectable even to
cosmological distances.
}
The millisecond durations of the recently-reported fast radio bursts
\citep[FRBs,][]{tsb+13} require a coherent emission process \citep{k13}; see also
\cite{lbmnc07} for a potential earlier example of the same
phenomenon.  The magnitude of various propagation effects is such that
they appear to have been generated at cosmological distances.  It is
not yet clear if they represent a previously unrecognized coherent
emission from a known source population or an entirely new class of
objects.
However, \cite{b-sbemc11} present a cautionary tale about the determination of
sources being at cosmological distances by relying solely on
propagation effects, and \cite{lsm13} present a model for FRBs in
which they originate from stars in the Galaxy.

Incoherent synchrotron emitting radio transients and variables span
the range from Galactic to cosmological distances, and many have
visible wavelength signatures.  Notable examples include active
galactic nuclei (AGN), gamma ray bursts (GRBs), supernovae (SNe), jetted
tidal disruption events (TDEs), X-ray binaries and microquasars, and
novae.  From the detection of radio wavelength synchrotron emission,
it is possible to obtain positional information (potentially with
milliarcsecond localization), energy information (e.g., velocity of
blastwave or relativistic outflow), beaming constraints, and
information on the density of the surrounding medium.  While radio
emission from these objects has been detected over a broad frequency
range, they tend to share a common evolution in frequency: generally
peaking earliest at the highest radio frequencies,
cascading to lower frequencies, and potentially remaining observable for
months to years at the lowest frequencies (\citealt*{wpms02}, see also
\citealt*{c82a,c82b,c98} and \citealt*{gps98}).
Strikingly, one of the key transient targets for
\hbox{LSST}, Type~Ia SNe, has never been detected at radio
wavelengths \citep{pvdwssm06}.

Recent interest in these kinds of objects tends to focus on short GRBs
and TDEs.  A subset of TDEs were recently discovered to launch relativistic jets,
likely in a similar manner to the radio jets in AGNs, 
allowing probes of the inner environment of a supermassive black hole \citep{bzp+12}.
Short GRBs are the leading candidates for an electromagnetic counterpart
to gravitational waves by the merging of a compact binary (e.g.,
neutron stars), but only a few short GRBs have been detected in the
radio to date \citep{fbm+13}.  Indeed,
possible radio counterparts to gravitational wave
events may include coherent bursts \citep[e.g.,][]{hl01,pp10} or
incoherent emitters \citep[e.g.,][]{bpc+05,sbk+06,np11,cf12} or both.

However, there are several challenges in observing gravitational wave
counterparts, either at visible or radio wavelengths.  The initial
source localizations could be poor \citep[tens to hundreds of square
degrees,][]{wc10,nsdh11,vma+12,nkg13,ligo+virgo13,kn13,rfrflfk13} with
the counterparts being short and faint at visible wavelengths, and
either very short ($\sim$ milliseconds) or occurring well after the
event (weeks to months) and faint in the radio \citep{mb13}.
\cite{fkobn12} find
that the sky appears to be far ``quieter,'' with fewer transients, at
centimeter wavelengths (radio) than at visible wavelengths, suggesting
that a radio counterpart may be the first identified counterpart to a
gravitational wave event, particularly if the radio counterpart is a
coherent burst.  These properties demand multi-wavelength programs,
with sensitive telescopes capable of covering large areas on the sky
(\S\S\ref{sec:survey}, Appendices~\ref{app:landscape}
and~\ref{app:lsst}), and a strong synergy exists between LSST and
radio surveys in identifying the electromagnetic counterpart at both
visible and radio wavelengths, and the information from both
wavelengths about the physics of the post-merger evolution will be
complementary.

Incoherent stellar emission is typically in the form of flares
generated as part of magnetic reconnection processes.  Thus, the
phenomenon is applicable to all classes of magnetically active stars.
Sensitive time-domain surveys, such as the \textit{Kepler} mission,
have been expanding the stellar types typically thought of for
flaring, particularly to~A stars \citep{b+12}.  M dwarfs have the best
constraints on their multi-wavelength characteristics (especially
radio and visible wavelength) and, with their large space densities,
are expected to be a significant contributor to the transient and
variable population that LSST will detect.  Because of the rapid
timescales involved, it is not feasible to make follow-up
observations, necessitating co-temporal coverage during coordinated
multi-wavelength observations.
Examples of processes in stellar coronae that are yet to be understood 
include the role of particle acceleration in the production of white stellar
flares as well as the upper limit on particle acceleration during large
stellar reconnection events.

Examples of Galactic coherent radio transients include neutron stars and
sub-stellar objects; there may also be coherent prompt emission
associated with violent events such GRBs \citep{uk00,sw02} or
gravitational wave events (e.g., binary neutron star mergers).  In
addition to their regular pulsations, radio pulsars have long been
known to have intermittent characteristics as well (e.g.,
``nulling'').  The extent and multi-wavelength aspects of their
intermittency is just beginning to be appreciated.
Radio pulsars have been found to turn off for days to weeks at a time
\citep[e.g., ][]{klojl06}, and a radio ``quiet'' mode in PSR~B0943$+$10
has been found to correlate with a large increase (more than double)
in its X-ray emission \citep{h+13}.
The extreme of intermittency is exhibited by rotating radio
transients (RRATs): neutron stars that produce only sporadic single radio
pulses \citep{mll+06}; some RRATs have been found to have X-ray
counterparts as well.

Although coherent radio emitters are not generally considered to have
visible wavelength counterparts, this situation may change in the LSST era.  For
instance, the annual LSST data releases, or the final (10~yr) data
release, may be deep enough to enable the detection of the visible wavelength
emission from neutron stars, which could be identified as such by
virtue of their colors and relatively high proper motions \citep[e.g.,][]{bcm93}.
One
strategy would be to confirm which, if any, known neutron stars are
detected by this approach.
Alternatively, any neutron star candidates
so identified within the LSST survey could be targeted with a deep
radio search.  In a similar vein, sub-stellar objects might either be
identified from LSST observations or be detected first in the radio then
identified in an LSST data release.  
As an example of the first method, nearby brown dwarf candidates could be 
identified in LSST by their red colors, then targeted for radio searches. 
As an example of the second method, searches of nearby stars
for electron cyclotron maser emission, characteristic of planets, could then
use LSST data to assess whether there have been intensity (transit) or astrometric 
variations of those stars.

\subsection{Discussion}\label{sec:transientdiscuss}

The workshop highlighted several key points regarding follow-up observations
of transients in the LSST era:
\begin{itemize}
\item%
X- and $\gamma$-ray telescopes, like those on \textit{Swift} and
\textit{Fermi}, have proven invaluable in the detection and
localization of various classes of transients.  In the case of
\textit{Swift}, it is capable of surveying roughly 1/3 of the sky
instantaneously and rapidly locating a burst to within a few
arcseconds.  During the LSST era, it is not clear that there will be a
comparable U.{}S.\ X- or $\gamma$-ray capability, though there may be
international capability.
\item%
An all-sky radio survey, similar to \hbox{NVSS}, but either deeper or 
higher frequency, or both, is invaluable for having a ``baseline
image'' and reference grid for when a transient is discovered.

\item%
Two-dimensional radio interferometers can provide and are essential
for obtaining precise ($< 1\arcsec$) positions for optical
cross-identifications and redshifts.  In constrast, a transient
detected by a single dish can be located only within the primary beam,
which is usually several arcminutes across and insufficient for
localization.

\item%
Using the knowledge that we have already acquired about the different
types of radio transients, and their rates and time scales
(Table~\ref{tab:transients}), is likely to be an important aspect of
the ability to classify LSST alerts and focus on the most
unusual or more interesting events.
\end{itemize}

More broadly, LSST is projected to produce as many as 1--2 million alerts
per night.  Formally, an ``alert'' is simply a notification that there
has been a difference between images, which may indicate a moving
object (e.g., asteroid), a variable, or a transient.  In order to
ensure that the necessary follow-up resources can be deployed, it is essential that appropriate classifier filters be
developed in order to reliably identify those alerts warranting
follow-up observations.  Specifically, event filters need to be
science-defined and have a carefully considered trade-off between
completeness and purity, or between identifying most of a population
(completeness) without being overwhelmed by false alarms (purity).  The community will need to develop and test
appropriate classification engines based on LSST alert data products,
a process that can begin now using on-going visible and near-infrared surveys.
Even with the most selective filters, the scientifically
``interesting'' event rate will overwhelm any plausible follow-up
resources.  Obtaining useful classifier filters will likely be a
multi-stage process, requiring a combination of the knowledge of
properties of known classes of transients and high quality pre-existing survey catalogs that provide multi-wavelength information.
Even after stringent filtering, attempting to follow up the
most interesting tip of the iceberg might still overwhelm available
resources.  Instead of triage by default, the community and the
observatories could consider defining \emph{which} types, if any, of
LSST alerts would warrant dedicated follow-up programs, as well as
defining how such follow-up programs would be conducted.

Given that the LSST Main Survey is expected to have a well-defined
cadence and schedule for observing fields, it may be possible to
combine this knowledge with knowledge of radio transient time scales
to do ``pre-observing'' or ``blind observing.''  For instance, if
there were to be a class of radio transients for which the radio
emission \emph{leads} the visible wavelength emission, it would be 
pragmatic 
to observe LSST fields
\emph{before} the LSST observations.  If the rate of such radio transients is
sufficiently high, a multi-wavelength study of a transient could be
conducted, even if the field in which it would be observed would not
be known \textit{a priori}.  A similar strategy has been used, to
great effect, in the observations of Type~Ia SNe with the
\textit{Hubble Space Telescope}.

\section{Surveys of the Sky}\label{sec:survey}


Radio continuum surveys provide an unobscured view of star formation and
accretion activity in galaxies out to the highest redshifts as well as a potentially powerful means of constraining cosmological
parameters.  Nearly all discrete radio sources are extragalactic and
are at cosmological distances $z \gtrsim 1$, so their distribution on
the sky is nearly isotropic. Surveys covering large solid angles are
needed to generate statistically useful samples of nearby ($z \lesssim
0.1$) radio galaxies or intrinsically rare objects (e.g., the most
luminous quasars), and to make sensitive cosmological tests (e.g.,
constraining dark energy via the Sachs-Wolfe effect).  Radio surveys
in the much smaller LSST ``Deep Drilling Fields'' ($\approx
10$~deg${}^2$ each, Appendix~\ref{app:lsst}) will be sufficient for
generating fair samples to study the evolution of AGNs and
star-forming galaxies.

The power that radio continuum surveys have to constrain the
cosmological model is becoming more apparent, mostly due to the
high-redshift tail of the radio source populations, which are
difficult to access in optical surveys.  To be of maximum value,
redshifts of the objects detected in radio continuum surveys are
required, but redshifts greater than unity will be difficult to obtain
even in the \hbox{SKA} era.
The SKA will measure the 21~cm \ion{H}{1}
line toward gas-rich galaxies, but the weakness of this line means
that individual galaxies will be difficult to detect in this line at
$z>1$, and it is unlikely to be useful for gas-poor galaxies.  Thus,
we will continue to rely on redshifts derived from optical and
near-infrared instruments.  The LSST's 6-band optical photometry will
provide accurate photometric redshifts to $z \sim 1.5$ (based on the
4000~\AA\ break spectral feature) and to $z > 2$ (based on the Lyman
break spectral feature); these photometric redshifts, particularly if
combined with complementary near-infrared photometry, will provide much
of the key data necessary to derive the evolution of activity in the
Universe, as traced by radio emission.  The possibility of obtaining a
large number of photometric redshifts for the $z < 2$ radio source
population with the LSST all-sky survey will allow the remaining
unidentified sources to be used for probing the largest scales at the
highest redshift \citep[e.g.,][]{rzb+12,c+12}.  Furthermore,
\hbox{LSST}'s multiple visits will probe photometric variability to
relatively faint magnitudes, providing additional information to aid
in determining which objects contain an active nucleus, and can help
classify AGN in the deep radio continuum surveys.

The K-correction of optically thin ($\alpha \sim -0.7$, $S_\nu \propto
\nu^\alpha$) synchrotron radio sources is so large that samples of
$z > 5$ sources may still be relatively small in number, even with
sensitive surveys at traditional frequencies ($\nu \approx
1.4\,\mathrm{GHz}$, but also see below).
Toward specific targets, deep, targeted observations have
been essential in obtaining precise positions, and the enhanced
sensitivity of the JVLA improves
this capability substantially.  Thus, it is now possible to use ultra-deep
1.4~GHz observations of well-studied fields, like the Chandra Deep
Field-North (CDF-N), which has both extensive spectroscopy (both
visible wavelength and CO) and X-ray imaging data, to find substantial
numbers of massive, star-forming galaxies to the highest redshifts.
At higher frequencies ($\nu \approx 5$--10~GHz), the
K-correction of free-free emission is more favorable.  Surveys at
these frequencies could have a high yield of intense star-forming
galaxies at $z > 5$, for which the (rest-frame) free-free emission
would be the direct result of the ionizing activity of massive stars.

Most radio sources stronger than $S \sim 1$~mJy at~1.4~GHz are powered
by supermassive black holes in AGNs, while star forming galaxies
increasingly dominate the population of fainter sources.  The radio
emission from a star forming galaxy is roughly co-extensive with the
star formation region, so the median angular size of faint sources is
no more than about~1\arcsec.  The 1.4~GHz source counts converge
rapidly below $S \sim 10\,\mu\mathrm{Jy}$: the flux density of a
normal galaxy at $z \sim 1$.  Consequently, instrumental confusion
declines sharply for surveys with better than 10\arcsec\ (FWHM)
angular resolution \citep{ccf+12}.  ``Natural'' confusion will never limit
sensitivity, unless there is a previously unrecognized population of
sources at microJanksy flux densities.

In addition to discrete sources, at the sensitivity levels of future
radio surveys, diffuse sources such as radio relics and halos in
clusters of galaxies could be detected.  Synchrotron ageing within
diffuse sources typically results in diffuse sources having steep
radio spectra, necessitating a different survey strategy.  Relatively
lower resolution surveys at frequencies lower than the ``standard''
of~1.4~GHz will be needed to maintain surface brightness sensitivity.
Extended or filamentary structures may also be traced by
cross-correlating the radio surveys with  the photometric galaxy density from
the LSST survey products.

Radio surveys in the LSST era should be designed to match both (1)~known
source properties and (2)~the planned LSST synoptic surveys covering
the whole southern hemisphere, plus up to 30 LSST Deep Drilling Fields covering approximately 10~deg$^2$ each.
Sensitive radio surveys should have detection limits well
below the median brightness temperature of star forming galaxies, $T_B \sim 1\,\mathrm{K}$ at~1.4~GHz.  For example, the 50~$\mu$Jy~beam${}^{-1}$ ($5\sigma$) detection limit in the 10\arcsec\ (FWHM) beam of the proposed Evolutionary Map of the Universe (EMU) survey of the southern
sky \citep{emu11} corresponds to $T_B \sim 0.3\,\mathrm{K}$. The sky density of LSST galaxies with $r < 27.7^\mathrm{m}$ is so high
(about one galaxy per~25 square arcseconds) that radio position errors
of order 0\farcs2 (rms) in each coordinate will be needed to make
complete and reliable position coincidence identifications.  Thermal
noise alone contributes errors of order 0\farcs1 FWHM for $5\sigma$
sources, so the deepest radio surveys in the Deep Drilling Fields may
have to be made with beams as small as approximately 2\arcsec\
\hbox{FWHM}.

SKA Precursor and pathfinder arrays (e.g., \hbox{ASKAP},
\hbox{MeerKAT}, \hbox{WSRT/APERTIF}, SKA Phase~1; see
Appendix~\ref{app:landscape}) may attain survey speeds over portions
of the radio spectrum (below~10~GHz, in most cases below~2~GHz)
superior to that of the \hbox{JVLA}, but the availability of the JVLA
opens the possibilities of starting long-term, large-area surveys now
that will be complementary to both future radio surveys and LSST
surveys.  There was considerable discussion of the benefits of such
JVLA surveys, such as earlier epochs for synoptic transient surveys,
higher frequency measurements, and follow up for the lower frequency
Precursor surveys, and radio counterparts for the visible and
near-infrared wavelength surveys that exist or are currently underway
(e.g., the \hbox{SDSS}, \hbox{PanSTARRS}, the Palomar Transient
Factory, Catalina Real Time Survey, \textit{Spitzer} Extragalactic
Representative Volume Survey). Additionally, data processing
and imaging techniques can be developed and tested that will be
important for enabling future arrays to realize their full
potential. Appendix~\ref{app:surveys} gives examples of potential
surveys that could be performed in the near future with the JVLA and
later with the SKA Phase~1, taking advantage of their new
capabilities, in preparation for the LSST era.

In the next few years, and complementing the timescale for any JVLA
surveys, the Low Frequency Array (LOFAR) will be conducting a series
of tiered surveys to explore the radio continuum sky
below~300~MHz. The key aims of these surveys are to trace the
evolution of activity with the deepest tiers and to constrain the
cosmological model through its large-area survey.  The frequencies at
which LOFAR operates also makes it highly efficient at discovering
steep-spectrum radio sources, such as radio halos and relics
associated with massive and likely merging clusters of galaxies, and 
high-redshift radio galaxies, which typically exhibit steep spectral
indices \citep[e.g.,][]{cmvh96,dvrm00,bretb98,cjrb07}.  
(However, \citealt*{rlmcs94} present a cautionary counter-example.)
The low radio-frequencies of the LOFAR surveys should
compensate for the large K-correction and could greatly increase
the sample of candidate radio sources at $z>5$ (and possibly into
the epoch of re-ionization), though spectroscopic or reliable photometric
redshifts will be required for conclusive identifications.

\section{Conclusions, Recommendations, and Future Developments}\label{sec:future}

While the workshop was structured around two science themes, there
were several topics discussed that were relevant to both.  These
topics included ``co-observing,'' probabilistic detection and
classification of sources, and the approach toward Deep Drilling
Fields.  Further, several questions were raised for which either
insufficient time for full discussion existed or no consensus was
reached.

``Co-observing'' was described as a technique for potentially more
efficient use of telescopes.  The LSST may broadcast its observing
schedule, allowing other observatories to know where the LSST will be
pointing.  In contrast to the current practice of coordinating
observations among multiple, independent telescopes, ``co-observing''
effectively removes a degree of freedom in the coordination, because
the LSST schedule can be treated as fixed and known.  Radio telescopes
could visit LSST fields with the appropriate lag to find certain
classes of sources.
However, while identified as a possible observing mode, a
specific science case or cases that would benefit explicitly from this
approach was not identified.

The use of machine learning or artificial intelligence techniques for
source detection and classification is not new in astronomy
\citep[e.g.,][]{wfd95,rgdp06}.  There was widespread agreement that
such techniques for the probabilistic detection and classification of
sources will be essential in the LSST era, if not well before.  It may not
be possible to determine with complete confidence whether two objects
at different wavelengths (e.g., visible and radio) that lie close
together on the sky are in fact the same object, particularly if the
flux density of one of the sources is near the detection threshold.
Similarly, in the early stages of some variable sources, it may not be
possible to distinguish between two or more classes of objects.

The locations of the Deep Drilling Fields yet to be selected were of
considerable interest.  
Many workshop participants found the locations
of the four selected Deep Drilling Fields\footnote{
\texttt{http://www.lsst.org/News/enews/deep-drilling-201202.html}
} (in the ``extragalactic sky'') to be well motivated.  However, there
was concern expressed that the distribution of strong radio sources in
or near a Deep Drilling Field may affect the dynamic range of the
radio images that could be obtained, thereby limiting the full extent
of multi-wavelength data that could be collected for them.
(This issue may also
affect the recently announced \textit{Hubble} Frontier
Fields.\footnote{
\texttt{http://www.stsci.edu/hst/campaigns/frontier-fields/}
})
Further discussion topics included the possibility of additional Deep Drilling Fields 
that have yet to be
selected, the need for at least one of these to be optimized for
Galactic targets, and the process by which input into the selection of
future Deep Drilling Fields could be provided\footnote{
For more information, see
\texttt{https://www.lsstcorp.org/content/whitepapers32012}.
}.

There was support expressed for inter-operability between software
tools, such as has been developed in the Virtual Observatory\footnote{
Internationally, Virtual Observatory activities are coordinated by the
International Virtal Observatory Alliance (IVOA)
} (VO)
context.  Given the multi-wavelength nature of many projects, as
stressed in this workshop, the ability to integrate results from
multiple surveys or observations will remain essential.  Further, for
time domain observations, the capability to distribute notices of
transients and obtain rapid response to those transients, if needed,
requires communication protocols and inter-operability.
More generally, the need for a better focus on topics such as
long-term data management and archiving was identified.  With data
volumes continuing to rise, even well before LSST becomes operational,
it is increasingly essential that projects take into account how their
data can be discovered, accessed, and analyzed.

There was considerable support for the notion that there is groundwork
that could be laid now, both in terms of time domain observations and
surveys.  While there exist mechanisms today for the rapid response of
alerts, it is not clear that these mechanisms have been
optimized, nor whether all avenues for rapid response have been
identified.  Similarly, there will be radio telescopes capable of
conducting deep observations or significant surveys well before the
LSST becomes operational.  Identifying the science case for such
surveys, and then executing them, would be important groundwork for
future comparisons with LSST images.

Once the LSST is operational, the resulting data set will render
unusual events commonplace, and the rarest of events observable.   The
challenge for the community is to determine how to use the LSST and its major sister facilities to determine how to optimize the follow-up observations that will be necessary to identify, classify, and characterize those events which represent new physics; clearly, given the size of the transient event pool, a set of stringent standards governing access to complementary facilities, both ground and space-based, will be essential to maximize the scientific output of the LSST, and to balance the ongoing operational missions of existing major facilities with time-critical follow-ups.

Meeting this challenge effectively will require establishing and
maintaining broad communications channels between the LSST project, 
its sister institutions, and within the community---at a level that
appears to have been relatively uncommon to date.   Issues which should be addressed might
include identifying and setting up appropriate information flows
between LSST and the complementary facilities that could be called
upon to provide follow-up or concurrent data, developing trigger
standards for such observations, making sure that these criteria
evolve to reflect ongoing experience as the LSST refines and ramps up
its observing programs, and making certain that sister observatories
develop and put in place processes to maintain the \textit{de facto}
community ownership of the LSST transient data pool.  As an example
specific to \hbox{NRAO}, how should observing time on the JVLA be
allocated to follow up particularly intriguing events?
Should time be granted based on standing trigger proposals (current
NRAO practice), or be restricted to first-come/first-serve telescope
time allocations only, or should NRAO consider setting aside a
fraction of its observing time for follow-up of LSST transient events
that satisfy certain community-established criteria, with the data
going immediately into the public domain, or some combination of all
of the above?  While this example is specific to the NRAO and the
\hbox{JVLA}, it may have a broader resonance with the entire nature of
``target of opportunity'' or ``rapid response'' proposals and the
process of time allocation at telescopes.

Finally, there was considerable discussion of future support.  In an
era of large surveys, with large numbers of people involved, how can
individual investigators obtain funding or be recognized for their
individual contributions?  At least in the U.{}S., there is unlikely
to be any program dedicated to LSST data analysis analogous to a
``guest observer'' program.\footnote{
There is no dedicated program within NSF for analysis of data from the
Atacama Large Millimeter/submillimeter Array (ALMA) or any other
ground-based NSF facility.
}  Among the approaches discussed, though
no concrete resolution was achieved, was the question of whether it is
possible to design a survey in a manner that would make it easier for
researchers involved in the survey to obtain funding.

\acknowledgements
We thank the National Radio Astronomy Observatory staff who provided
logistical support for this workshop, particularly K.~Ransom and
C.~Hunsinger.  
Part of this research was carried
out at the Jet Propulsion Laboratory, California Institute of
Technology, under a contract with the National Aeronautics and Space Administration.
Basic research in radio astronomy at the Naval Research Laboratory is supported by~6.1 Base funding.
The National Radio Astronomy Observatory is a facility of the National
Science Foundation operated under cooperative agreement by Associated
Universities, Inc.
Manuscript \textcopyright~2013.  All rights reserved.

\clearpage

\appendix

\section{Radio Astronomy Landscape}\label{app:landscape}

Tables~\ref{tab:singledish}, \ref{tab:meter},
and~\ref{tab:centimeter} highlight radio telescopes either currently
operational or in construction and likely to be operational in the
LSST era.  These tables may not be complete, and, particularly for
telescopes still under construction or in commissioning, various aspects of their
capabilities may change.
\cite{dtmmlc13} have conducted  a similar comparison.  

\begin{deluxetable}{lcccccc}
\rotate
\tablecaption{Single Dish Radio Telescopes\label{tab:singledish}}
\tabletypesize{\footnotesize}
\tablewidth{0pt}
\tablehead{%
 \colhead{Parameter} 
	& \colhead{Arecibo} & \colhead{Effelsberg}
	& \colhead{GBT\tablenotemark{a}} & \colhead{Parkes} & \colhead{\textit{FAST}}}
\startdata
Aperture              & 300~m                  & 100~m
	& 100~m    & 64~m                  & 500~m  \\
Frequency Range (GHz) & 0.3--10                & 0.4--86
	& 0.2--100 & 0.7, 1.2--1.8, 2.3, 4.8  & 0.07--3 \\
		      & (\hbox{ALFA}: 1.4~GHz) &                
	&          & 8.5, 12--15, 16--26 &   \\
  		      &                        &
	&          & (multi-beam: 1.4~GHz) &   \\

Resolution (@ 1.4~GHz) & 3\farcm4 	& 8\arcmin
	& 7\arcmin & 14\farcm8                 & 2\farcm9 \\
                       & (\hbox{ALFA}: 7 $\times$ 3\farcm4) & 
	&          & (multi-beam: 14\farcm2) \\

Sensitivity\tablenotemark{b}\,($\mu$Jy~MHz${}^{-1/2}$~hr${}^{-1/2}$) &
	45 & 330 & 112 & 500 & 1050 \\
Survey Speed\tablenotemark{c}\,(deg${}^2$~hr${}^{-1}$) &
	1.8               & 0.5  & 1.1  & 0.76 & 2.5             \\
                                                      &
	(\hbox{ALFA}: 9.0) &      &      & (multi-beam: 4.6) \\

Location             & USA     & Germany                  
	& USA   & Australia & China \\
\enddata
\tablenotetext{a}{Listed are the current operational parameters of the
Green Bank Telescope.  As a result of the Portfolio Review conducted
by the U.{}S.\ National Science Foundation, the operational model for
the GBT will change in or before~2017.}
\tablenotetext{b}{The quoted values are typically at frequencies
near~1.4~GHz, are intended to be illustrative of the relative
sensitivity of the various telescopes, and assume a 1~MHz processed bandwidth
and 1~hr integration time, with the sensitivity scaling as
$[(\Delta\nu/1\,\mathrm{MHz})(\Delta t/1\,\mathrm{hr})]^{-1/2}$.  In
practice, the sensitivity is a function of frequency and all of these telescopes will be classically confusion
limited in a much shorter duration than 1~hr, but surveys for spectral
lines or time domain observations can approach these sensitivity
levels.}
\tablenotetext{c}{The quoted values are at frequencies near~1.4~GHz
and are intended to be illustrative of the relative survey speed of
the telescopes.  The quoted values are the area (in square degrees)
that can be surveyed to a flux density~$\Delta S$ of~0.1~mJy (rms) in~1~hr, assuming a
processed bandwidth~$\Delta\nu$ of~100~MHz.  The survey speed scales
as $\Delta\nu(\Delta S)^2$.}
\tablecomments{
FAST = Five hundred metre Aperture Spherical Telescope; GBT = Green
Bank Telescope.  Telescopes listed in \textit{italic style} are under
construction, and the values listed should be considered notional.}
\end{deluxetable}

\begin{deluxetable}{lccccc}
\rotate
\tablecaption{Meter-wavelength Interferometers\label{tab:meter}}
\tablewidth{0pt}
\tabletypesize{\footnotesize}
\tablehead{%
 \colhead{Parameter} & \colhead{LOFAR} & \colhead{LWA1} &
\colhead{\textit{LWA-OVRO}} & \colhead{MWA} & \colhead{PAPER}}
\startdata
Frequency range (MHz) & 10--90, 110--250      & 10--88            & 10--88
	& 80--300 & 110--180 \\
Angular Resolution    & 10\arcsec\ (@ 60~MHz) & 2\fdg5 (@ 80~MHz) & 1\fdg3 (@ 80~MHz)
	& 3\arcmin\ (@ 150~MHz) & 0.5\arcdeg\tablenotemark{a} \\
Field of View         & 9.77\arcdeg (@ 60~MHz)\tablenotemark{b} & 2\fdg5 (@ 80~MHz) & 135\arcdeg\ (@ 80~MHz)
	& 25\arcdeg\ (@ 150~MHz) & 60\arcdeg \\ 
Location              & The Netherlands, Europe & USA & USA &
	Australia & \hbox{USA}, South Africa \\
\enddata
\tablecomments{LOFAR = Low Frequency Array; LWA = Long Wavelength
Array; MWA = Murchison Wide-field Array; PAPER = Precision Array to
Probe the Epoch of Reionization.  Telescopes listed in \textit{italic style} are under
construction, and the values listed should be considered notional.}
\tablenotetext{a}{The listed value is appropriate when PAPER is
configured in an imaging configuration.}
\tablenotetext{b}{The field of view for LOFAR depends upon the antenna
stations used.  The value listed is the maximal value, and it scales
approximately with frequency.}
\end{deluxetable}

\begin{deluxetable}{lcccccccccc}
\rotate
\tablecaption{Centimeter-wavelength
	Interferometers\label{tab:centimeter}}
\tabletypesize{\scriptsize}
\tablewidth{\textheight}
\tablehead{%
 \colhead{Parameter} 
	& \colhead{\textit{ASKAP}}
	& \colhead{ATCA}
	& \colhead{\textit{CHIME}\tablenotemark{a}}
	& \colhead{eMERLIN} 
	& \colhead{EVN\tablenotemark{b}}
}
\startdata
Frequency range (GHz) 
	& 0.7--1.7                                      
	& 1.1--3.1, 4.0--10.8, 16--25,                  
	& 0.4--0.8                                      
	& 1.3--1.8, 4--8, 22--24                        
	& 0.33, 0.61, 1--1.67, 2.5,                     
\\
	&                                               
	& 30--50, 83.5--106                             
	&                                               
	&                                               
	& 5--6, 8.5, 22, 43			        
\\
Angular Resolution (\arcsec, at~1.4~GHz)
	& 10                                            
	& 6.8 						
	& (800) 					
	& 0.15                                          
	& 0.004                                         
\\		
Field of View (deg${}^2$, at~1.4~GHz)			
	& 30                                            
	& 0.21						
	& ($180\arcdeg\times1.3\arcdeg$)		
	& 0.13 						
	& \ldots                                        
\\
Bandwidth (MHz)
	& 300						
	& 2000						
	& 400						
	& 400						
	& 256						
\\
Point-source Sensitivity\tablenotemark{e}\,\,($\mu$Jy~hr${}^{-1/2}$)
	& 47                                            
	& 14						
	& (16) 						
	& 27 						
	& 220 						
\\
Survey Speed\tablenotemark{f}\,(deg${}^2$~hr${}^{-1}$ to~0.1~mJy rms)
	& 348                                           
	& 11 						
	& (8900)					
	& \ldots\tablenotemark{g}			
	& \ldots\tablenotemark{g}                       
\\
Location    
	& Australia                                     
	& Australia					
	& Canada					
	& England					
	& Europe					
\\
\noalign{\hrule}\\
\noalign{\hrule}
	& \colhead{GMRT} 
	& \colhead{\textit{MeerKAT}} 
	& \colhead{JVLA} 
	& \colhead{VLBA\tablenotemark{c}}
	& \colhead{WSRT\tablenotemark{d}} \\
Frequency range (GHz) 
	& 0.15, 0.23, 0.33, 	                        
	& 0.9--1.67, 8--14.5                            
	& 1--50                                         
	& 0.33--86					
	& 0.25--0.46, 0.59--0.61, 0.7--1.2,   		
\\
	& 0.61, 1.4                                     
	&                                               
	&                                               
	&                                 		
	& 1.15--1.7, 2.3, 4.8, 8.4	                
\\
	&                                               
	&                                               
	&                                               
	&                                 		
	& (\hbox{APERTIF}: 1.4~GHz)                    	
\\
Angular Resolution (\arcsec, at~1.4~GHz)
	& 2                                             
	& 5.9 						
	& 1.2 						
	& 0.005 					
	& 12                     			
\\		
Field of View (deg${}^2$, at~1.4~GHz)			
	& 0.05                                          
	& 0.56						
	& 0.16						
	& \ldots					
	& 0.16 (\hbox{APERTIF}: 8)			
\\
Bandwidth (MHz)
	& 120						
	& 500						
	& 600						
	& 256						
	& 144						
\\
Point-source Sensitivity\tablenotemark{e}\,\,($\mu$Jy~hr${}^{-1/2}$)
	& 13                                            
	& 10 						
	& 7.6 						
	& 21                                            
	& 28                                            
\\
Survey Speed\tablenotemark{f}\,(deg${}^2$~hr${}^{-1}$ to~0.1~mJy rms)
	& 5.1	                                      
	& 75	                  			
	& 21	                                        
	& \ldots\tablenotemark{g}		        
	& 4.1 (\hbox{APERTIF}: 116)			
\\
Location    
	& India 					
	& South Africa					
	& USA						
	& USA						
	& The Netherlands				
\enddata
\tablenotetext{a}{CHIME is not planned to operate at~1.4~GHz.  We list
	values appropriate for its highest frequency, 800~MHz,
	corresponding to a wavelength~$\lambda$37.5cm.}
\tablenotetext{b}{As an inhomogeneous array, not all antennas within
	the EVN operate at all of the frequencies listed.  A majority
	operate at or near~1.4~GHz, which is used as an illustrative
	frequency for many of the other parameters.}
\tablenotetext{c}{Listed are the current operational parameters of the
	Very Long Baseline Array.  As a result of the Portfolio Review
	conducted by the U.{}S.\ National Science Foundation, the
	operational model for the VLBA will change in or before~2017.}
\tablenotetext{d}{The standard operational mode for the Westerbork
	Synthesis Radio Telescope is a 12~hr observation.}
\tablenotetext{e}{The quoted values are at frequencies
	near~1.4~GHz and are intended to be illustrative of the relative
	sensitivities of the various telescopes.  The 
	sensitivity scales as $[(\Delta\nu)(\Delta
	t/1\,\mathrm{hr})]^{-1/2}$.}
\tablenotetext{f}{The quoted values are at frequencies near~1.4~GHz
	and are intended to be illustrative of the relative survey
	speeds of the telescopes.  The quoted values are the area (in
	square degrees) that can be surveyed to a flux density~$\Delta
	S$ of~0.1~mJy (rms) in~1~hr.  The survey speed scales as
	$\Delta\nu(\Delta S)^2$.}
\tablenotetext{g}{These high angular resolution instruments were not
	intended to conduct blind surveys over the entire field of
	view.  However, recent correlator developments do enable the
	simultaneous observations of many, potentially hundreds, of the radio sources within
	the field of view, allowing ``targeted surveys.''}
\tablecomments{%
ASKAP = Australian Square Kilometre Array Pathfinder; ATCA = Australia
Telescope Compact Array; CHIME = Canadian H\,\textsc{i} Mapping
Experiment; eMERLIN = enhanced Multi-Element Radio Linked
Interferometric Network; EVN = European Very Long Baseline Interferometry Network;
GMRT = Giant Metrewave Radio Telescope; MeerKAT = Karoo Array
Telescope; JVLA = Karl G.~Jansky Very Large Array; VLBA = Very Long Baseline Array;
WSRT = Westerbork Synthesis Radio Telescope. Telescopes in
\textit{italic style} are under construction or design, and the values
listed should be taken as notional.}
\end{deluxetable}

We stress that these values should be taken as illustrative of the
relative performance, but there are many factors that affect the
ultimate performance for a given observing program.  
The observatories that operate most of the
telescopes listed maintain up-to-date status documents that should be
consulted for the most recent values.

\section{LSST Data Products}\label{app:lsst}

\newcommand{\B}[1]{{#1}}

LSST will be a large, wide-field ground-based optical telescope system
designed to obtain multiple images covering the sky that is visible
from Cerro Pach\'{o}n in Northern Chile. The current baseline design,
with an 8.4~m (6.7~m effective) primary mirror, a 9.6~deg$^2$ field of
view, and a 3.2 Gigapixel camera, will allow about 10\,000~deg${}^2$
of sky to be covered every three to four nights using pairs  of 15~second exposures, with typical 5$\sigma$ depth for point sources of $r\sim24.5$ (AB). The system is designed to yield high image quality as well as superb astrometric  and photometric accuracy. The \B{total} survey area will include $\sim$30\,000~deg$^2$ with $\delta<+34.5^\circ$, and will be imaged multiple times in six bands, $ugrizy$, covering the wavelength range 320--1050~nm.

The project is scheduled to begin the regular survey operations at the
start of next decade. About 90\% of the observing time will be devoted
to a deep-wide-fast survey mode (``Main Survey'') which will
\B{uniformly} observe an 18\,000~deg$^2$ region about 1000 times
(summed over all six bands) during the anticipated 10~years of
operations, and yield a coadded map to $r\sim27.5$. These data will
result in catalogs including over $38$ billion stars and galaxies,
that will serve the majority of the primary science programs. The
remaining 10\% of the observing time will be allocated to special
projects such as a Very Deep and Fast time domain
survey.\footnote{Informally known as ``Deep Drilling Fields,''
notionally expected to cover approximately 10~deg${}^2$ each.}

LSST Data Management will perform, in an automated fashion, two types of image analyses resulting in two levels (classes) of Data Products:
\begin{enumerate}
\item Analysis of difference images, with the goal of detecting and
characterizing astrophysical phenomena revealed by their
time-dependent nature. The detection of supernovae superimposed on
bright extended galaxies is an example of this analysis. The
processing will be done on nightly or daily basis and result in {\bf
Level~1} data products. Level~1 products will include difference
images, catalogs of sources detected in difference images,
astrophysical objects to which sources in the difference images are
associated, and catalogs of orbits of identified Solar System
objects. The catalogs will be entered into the {\bf Level~1 Database}
and made available in near real time. Notifications (``alerts'') about
differences between images will be issued using community-accepted standards.
The primary results of analysis of difference images---discovered and
characterized sources---will generally be broadcast as {\em event
alerts} within 60~seconds of end-of-visit acquisition of a field,
allowing for rapid follow up, lest interesting information about a
source be lost.

\item Analysis of direct images, with the goal of detecting and characterizing astrophysical objects. Detection of faint galaxies on deep co-adds and their subsequent characterization is an example of this analysis. The results are {\bf Level~2} data products. These products, generated and released annually, will include the single-epoch images, deep co-adds, catalogs of characterized objects (detected on deep co-adds as well as individual visits), sources (detections and measurements on individual visits), and {\em forced sources} (constrained measurement of flux on individual visits using known source positions). It will also include fully reprocessed Level~1 data products. In contrast to the Level~1 catalog, which is updated in real-time, the data releases will be static and will not change after release.
The analysis of science (direct) images is less time sensitive, and will be done as a part of annual data release process.

\item 
Recognizing the diversity of astronomical community needs, as well as the need for specialized processing not part of the automatically generated Level~1 and~2 products, LSST plans to make its software and APIs available for community use, and to devote 10\% of its data management system capabilities to enabling the creation, use, and federation of {\bf Level~3} (user-created) data products. Level~3 capabilities will help bridge the gap between LSST Data Products and the science envisioned for \hbox{LSST}, and enable science cases that greatly benefit from co-location of user processing and/or data within the LSST Archive Center.
\end{enumerate}

\clearpage

\section{Prospective Radio Sky Surveys\label{app:surveys}}

\subsection{Karl G.~Jansky Very Large Array}\label{sec:jvla}

Nearly 20~years ago, the pioneering FIRST and NVSS radio surveys used
the Very Large Array for the first time to carry out large wide area,
community accessible surveys and data products. With the newly
upgraded JVLA construction complete and
full science operations underway, the next generation of  large JVLA
sky surveys is now possible. 
With expanded radio frequency and
correlator bandwidth, the JVLA is 7 to nearly 100 (at~1--2 and
40--48~GHz, respectively) times faster for continuum and line search
studies than the classic \hbox{VLA}.  
An effort is now underway to plan and implement the next large VLA Sky
Survey, the spiritual successor to FIRST and \hbox{NVSS}. 

\begin{deluxetable}{llcccc}
\tablecaption{Survey Speed of the JVLA\label{tab:ss}}
\tabletypesize{\small}
\tablehead{%
\colhead{Frequency} & \colhead{Bandwidth}
	& \colhead{$\Delta t_{\mathrm{int}}$} & \colhead{Field of View} 
	& \colhead{Survey Speed}              & \colhead{Mapping Rate} \\
\colhead{(RF Band)} & \colhead{(Mode)}
	& \colhead{(s)}                       & \colhead{(\arcmin)}
	& \colhead{(deg$^2$~hr$^{-1}$)}   & \colhead{(deg~min$^{-1}$)}
}
\startdata

1--2~GHz (L)      & 0.6~GHz (8~bit)  & 37 & 30    & 13.9 & 0.65 \\
2--4~GHz (S)      & 1.5~GHz (8~bit)  &  7.7 & 15    & 16.5 & 1.56 \\
4--8~GHz (C)      & 1.8~GHz (8~bit)  &  5.5 &  7.5  &  5.7 & 1.08 \\
                  & 3.03~GHz (3~bit) &  4.4 &  7.5  &  7.2 & 1.36 \\
8--12~GHz (X)     & 3.50~GHz (3~bit) &  3.9 &  4.5  &  2.9 & 0.93 \\
12--18~GHz (Ku)   & 5.25~GHz (3~bit) &  3.5 &  3    &  1.4 & 0.68 \\
18--26.5~GHz (K)  & 7~GHz (3~bit)    &  7.1 &  2.05 &  0.3 & 0.23 \\
26.5--40~GHz (Ka) & 7~GHz (3~bit)    &  9.7 &  1.45 &  0.1 & 0.12 \\
40--50~GHz (Q)    & 7~GHz (3~bit)    & 50.4 &  1    &  0.01 & 0.02 \\

\enddata
\tablecomments{The field of view is defined to be the full width, half
maximum of the antenna power pattern, calculated at band center for
the effective  interference-free frequency ``width'' for the chosen 
sampling mode (8-bit or 3-bit). The ``Mapping Rate'' is the on-the-fly scanning rate (deg~min$^{-1}$) needed to reach this depth.}
\end{deluxetable}

Table~\ref{tab:ss} summarizes the \emph{survey speed} of the
\hbox{JVLA}, defined as the area per unit time (deg${}^2$~hr${}^{-1}$)
that can be imaged in the continuum to a point source ($1\sigma$
thermal noise) sensitivity of~100~$\mu$Jy (assuming natural weighting).
Examples of possible JVLA Sky Surveys include:
\begin{itemize}
\item An NVSS-style survey ($\sim 30\,000$~deg${}^2$) at~2--4~GHz---This
survey would require about 1800~hr of integration time to reach
100~$\mu$Jy in a single epoch and could be conducted so as to provide both high angular resolution and reasonable
surface brightness sensitivity (e.g., using A and~C configurations to
provide an angular resolution approaching 1\arcsec\ while maintaining
sensitivity to scales as large as 90\arcsec).
For reference, the NVSS \citep{nvss} program took about~2700~hr to reach a level of~450~$\mu$Jy (rms) at~1.4~GHz with an angular resolution of~45\arcsec\ (D configuration). 

\item A FIRST-style survey ($\sim 10\,000$~deg${}^2$) at~4--8~GHz---This
survey would require about 1400~hr of integration time to reach
100~$\mu$Jy in a single epoch and could be repeated every other
configuration cycle (32~months).  The angular resolution could
approach 0\farcs5 (A-configuration, natural weighting). 
For reference, the 1.4~GHz FIRST survey \citep{first} took
about~3200~hr to reach a level of~150~$\mu$Jy (rms) with an angular resolution of~5\arcsec\ (B
configuration).

\item A medium-deep 1000~deg${}^2$ synoptic survey at~8--12~GHz---This
survey  would require about~400~hr of integration time to  reach
100~$\mu$Jy and could be repeated every 16~months (corresponding to the
time it takes the JVLA to cycle through its  configurations).  The
angular resolution could approach 0\farcs3 (A-configuration, natural
weighting).  This combination of angular resolution and sensitivity
would be sufficient to resolve some gravitational lenses while the
repetition every configuration cycle would allow identification and
monitoring of long duration transients such as TDEs
(\S\ref{sec:radiotransients}).

\item A 1000~deg${}^2$ survey at~12--18~GHz---This survey would
require 700~hr of integration time and could have an angular
resolution approaching 0\farcs2 (A-configuration),
which might be particularly attractive in the crowded Galactic plane.

\item The previously detailed 1000~deg${}^2$ synoptic surveys could,
at little additional effort, include a lower frequency (1--2~GHz or~2--4~GHz) contemporaneous observation.

\item A repeated coordinated campaign to map and monitor the
accessible LSST Deep Drilling Fields more deeply in a number of bands could be carried out separately or as part of a larger multi-tiered survey program.
\end{itemize}
Although the above examples are directed at continuum surveys, the
wide spectral bandwidths available would also enable simultaneous line
surveys of (for example) \ion{H}{1}, redshifted \hbox{CO}, maser
emission, or recombination lines.  Blind CO surveys may be
particularly profitable, especially if made in well studied regions
such as the COSMOS field. In addition to the main suite of receivers
covering 1--50~GHz, new new low-frequency receivers covering
230--470~MHz (P~band) and 58--84~MHz (``4-band'' or VHF band) are
being commissioned for the \hbox{JVLA}.
Furthermore, a commensal low frequency observing system (low-band 
observatory, LOBO) for the JVLA is under discussion. 
A limited duration, 10-antenna pathfinder (VLA Low Frequency
Ionosphere and Transient Experiment, VLITE) is funded and under
development in a partnership between NRAO and the U.{}S.\ Naval
Research Laboratory.  If successful and expanded to \hbox{LOBO}, low-frequency
observing could be carried out simultaneously with any of the previous examples.

\subsection{Square Kilometer Array Phase~1}\label{ref:ska1}

The key requirements for the SKA Phase~1 to push beyond the Pathfinder instruments are the increase in sensitivity (SKA1\_Mid) and survey speed (SKA1\_Survey), which allow these surveys to be conducted within 2 years of observing time (similar to ASKAP key projects) or less. 
The proposed longest baselines of~50~km for SKA1\_Survey and~200~km for SKA1\_Mid should be sufficient to mitigate instrumental confusion.  
The SKA Organisation released a baseline design for Phase~1 in~2013 March, describing the basic 3-telescope (SKA\_Mid, SKA1\_Survey and SKA1\_Low) model. Although not the final design, the baseline can be used to identify potential continuum and HI surveys with SKA Phase 1:

\begin{description}
\item[Ultra-deep continuum survey with SKA1\_Mid]%
An ultra-deep continuum survey would probe galaxy evolution and
star-formation over cosmic time, detect AGN to the Epoch of
Reionization, and cover a large enough volume to probe clustering of
massive galaxies \citep[e.g.,][]{saa+13}.  An ultra-deep survey
conducted at~1.4~GHz should reach a 5$\sigma$ detection limit of at
least 250~nJy and cover at least 6~deg$^2$.  A resolution of
$\sim1$\arcsec\ will mitigate confusion. An imaging dynamic range of
$4 \times 10^6$ (66~dB) is required, as one 0.1~Jy source is expected
in each $\sim$1 deg field of view.

\item[Medium-deep continuum survey with SKA1\_Mid]%
A medium-deep survey would probe dark matter via weak lensing. Such a
survey must cover at least 5000~deg$^2$ to be competitive with the
Dark Energy Survey\footnote{
\texttt{http://www.darkenergysurvey.org/}
}
\citep[\hbox{DES}, ][]{des10}, which is expected to be completed by~2017.  The SKA1 survey should achieve at least 0.5~$\mu$Jy rms at~1.4~GHz with 0.5\arcsec\ resolution, resulting in about 7 resolved radio sources per square arcminute with a median redshift $z \sim 1.6$ (D.~Bacon~2013, private communication, using S$^3$ simulations of \citealt*{s3}).

\item[All-Sky Continuum Survey with SKA1\_Survey]%
An all-sky radio continuum survey would probe cosmology via two
important tests: the integrated Sachs-Wolfe effect and cosmic
magnification. This survey should cover at least 20,000~deg$^2$
to~1~$\mu$Jy (rms) at~1.4~GHz. A resolution better than 2.5\arcsec\ is
required to mitigate confusion. An imaging dynamic range of order $10^7$ (70~dB) is required, as one 10~Jy source is expected in each $\sim$20~deg${}^2$ field of view.

\item[All-Sky H\,{I} Survey with SKA1\_Survey]%
An all-sky \ion{H}{1} survey could be performed commensurately with
the all-sky continuum survey in order to probe the evolution of
\ion{H}{1} and its role in galaxy evolution. A large-area radio
spectroscopic survey could also be used to constrain cosmological
parameters via the detection of baryonic acoustic oscillations. An
all-sky \ion{H}{1} survey with SKA1\_Survey could reach 85~$\mu$Jy
(rms) in~0.1~MHz channels over 20,000~deg$^2$, yielding roughly 7
million galaxy detections.
\end{description}

\clearpage

\section{Participants}\label{app:who}

Table~\ref{tab:participate} lists the participants in the workshop.

\begin{deluxetable}{ll}
\tablecaption{Workshop Participants\label{tab:participate}}
\tablewidth{0pc}
\tabletypesize{\small}
\tablehead{%
 \colhead{Name} & \colhead{Institution}}
 \startdata
  Andrew Baker &  Rutgers, the State University of New Jersey \\
  Amy Barger &  University of Wisconsin \\
  Tim Bastian &  NRAO \\
  Tony Beasley &  NRAO \\
  Niel Brandt &  Penn State University \\
  Dario Carbone &  University of Amsterdam \\
  Patti Carroll &  University of Washington \\
  Tzu-Ching Chang &  ASIAA \\
  Shami Chatterjee &  Cornell University \\
  Tracy Clarke &  Naval Research Laboratory \\
  Jim Condon &  NRAO \\
  Anca Constantin &  James Madison University \\
  Steve Croft &  UC Berkeley / University of Wisconsin-Milwaukee \\
  Sean Cutchin &  NRC / NRL \\
  Bob Dickman &  NRAO \\
  Sean Dougherty &  NRC \\
  Michael Garrett &  ASTRON \\
  Jason Glenn &  University of Colorado \\
  Gregg Hallinan &  Caltech \\
  Robert Hanisch &  STScI/VAO \\
  Assaf Horesh &  Caltech \\
  Minh Huynh &  University of Western Australia \\
  Scott Hyman &  Sweet Briar College \\
  Matt Jarvis &  Oxford University \\
  Dayton Jones &  JPL/Caltech \\
  Mario Juric &  LSST \\
  Namir Kassim &  Naval Research Laboratory \\
  Ken Kellermann &  NRAO \\
  Brian Kent &  NRAO \\
  Amy Kimball &  NRAO \\
  Mark Lacy &  NRAO \\
  Cornelia Lang &  University of Iowa \\
  Casey Law &  UC Berkeley \\
  Joseph Lazio &  Jet Propulsion Laboratory, California Institute of Technology \\
  Dana Lehr &  National Science Foundation \\
  Sam Lindsay &  University of Hertfordshire \\
  Brian Mason &  NRAO \\
  Walter Max-Moerbeck &  NRAO \\
  Mark McKinnon &  NRAO \\
  Kunal Mooley &  California Institute of Technology \\
  Tony Mroczkowski &  Caltech / JPL \\
  Steven Myers &  NRAO \\
  Bojan Nikolic &  University of Cambridge \\
  Rachel Osten &  STScI \\
  Joshua Peek &  Columbia University \\
  Phil Puxley &  National Science Foundation \\
  Peter Quinn &  International Centre for Radio Astronomy Research (ICRAR) \\
  Bart Scheers &  \hbox{UvA} / CWI \\
  Nigel Sharp &  NSF \\
  Ohad Shemmer &  University of North Texas \\
  Albert Stebbins &  Fermilab \\
  Tony Tyson &  UC Davis \\
  Paul Vanden Bout &  NRAO \\
  Lucianne Walkowicz &  Princeton University \\
  Peter Williams &  Harvard \\
  B.~Ashley Zauderer &  Harvard \\
\enddata
\end{deluxetable}

\clearpage

\clearpage

\end{document}